# A knowledge-assisted visual malware analysis system: Design, validation, and reflection of KAMAS

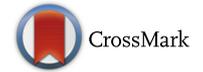

Markus Wagner [a,b,*], Alexander Rind [a,b], Niklas Thür [a], Wolfgang Aigner [a,b]

[a] *St. Pölten University of Applied Sciences, St. Pölten, Austria*
[b] *Vienna University of Technology, Vienna, Austria*

ABSTRACT

IT-security experts engage in behavior-based malware analysis in order to learn about previously unknown samples of malicious software (malware) or malware families. For this, they need to find and categorize suspicious patterns from large collections of execution traces. Currently available systems do not meet the analysts' needs which are described as: visual access suitable for complex data structures, visual representations appropriate for IT-security experts, provision of workflow-specific interaction techniques, and the ability to externalize knowledge in the form of rules to ease the analysis process and to share with colleagues. To close this gap, we designed and developed KAMAS, a knowledge-assisted visualization system for behavior-based malware analysis. This paper is a design study that describes the design, implementation, and evaluation of the prototype. We report on the validation of KAMAS with expert reviews, a user study with domain experts and focus group meetings with analysts from industry. Additionally, we reflect on the acquired insights of the design study and discuss the advantages and disadvantages of the applied visualization methods. An interesting finding is that the arc-diagram was one of the preferred visualization techniques during the design phase but did not provide the expected benefits for finding patterns. In contrast, the seemingly simple looking connection line was described as supportive in finding the link between the rule overview table and the rule detail table which are playing a central role for the analysis in KAMAS.

## 1. Introduction

Malware (malicious software) is undoubtedly one of today's greatest threats to the Confidentiality/Integrity/Availability (CIA) triangle of information security (Stoneburner et al., 2002). It has become a common tool in digital theft, corporate and national espionage, spam distribution and attacks on infrastructure availability. When security professionals analyze malware in a real world setting, they have large volumes of complex and heterogeneous data at their disposal. In behavior-based malware analysis, they explore traces of malware execution in the form of system call sequences (rules) and frequently occurring subsequences of potentially malicious code (Egele et al., 2008). Their workflow involves the tasks of selecting different rules, categorizing them, and storing them in a database as well as manual adaption and/or tuning of found rules (Wagner et al., 2014). In addition to challenging analysis methods, "implicit knowledge" (Chen et al., 2009) or "tacit knowledge" (Wang et al., 2009) about the data, the domain experience or prior experience are often required to make sense of the data and not become overwhelmed. By externalizing some of the domain experts' implicit knowledge, it can be made available as explicit knowledge and stored in a knowledge database (KDB) (Chen et al., 2009).

---

* *Corresponding author.*
  *E-mail addresses:* markus.wagner@fhstp.ac.at (M. Wagner), alexander.rind@fhstp.ac.at (A. Rind), niklas.thuer@fhstp.ac.at (N. Thür), wolfgang.aigner@fhstp.ac.at (W. Aigner).

In this paper, we present a visualization design study in the context of malware analysis. Visualization tools "provide visual representations of datasets designed to help people carry out tasks more effectively" (Munzner, 2014). Thus, they are particularly useful "to augment human capabilities rather than replace people with computational decision-making methods" (Munzner, 2014). Malware analysis lends itself very well to visualization, because the experience of analysts plays a central role in reconstructing the obfuscated behavior of malware. Users' needs in the context of behavior-based malware analysis were analyzed systematically in previous work (Wagner et al., 2014) and are summarized in Section 2. Currently, there are no interactive visualization tools available which cover all the needs of the malware analysis experts. To close this gap, we developed a novel Knowledge-Assisted Visual Malware Analysis System (KAMAS) (see Fig. 1). Additionally, we demonstrate how the visualization can benefit from explicit domain knowledge.

In order to achieve the best possible results, we followed the paradigm of problem-oriented research, i.e., collaborating with real users to solve their tasks (Sedlmair et al., 2012). Therefore, we worked in accordance with the *nested model for visualization design and validation* by Munzner (Munzner, 2009), which divides visualization design into four levels combined with appropriate validation methods (see Section 2). Specifically, we focused on the third and fourth level of Munzner's model (Munzner, 2009) (third level: visual encoding and interaction design, fourth level: algorithm design). Moreover, we evaluated the prototype's usability in relation to the interaction metaphors and knowledge representation needed by IT-security experts. Thus, the main contributions of our research are:

- We present the concept and implementation of KAMAS as systematically designed, developed and evaluated instantiation of an interactive visual method for handling large amounts of complex data in behavior-based malware analysis.
- We show that applying knowledge-assisted visual analytics methods allows domain experts to externalize their implicit knowledge and profit from this explicit knowledge in their analysis workflow.
- To provide evidence for the effectiveness of the developed methods, we provide a rigorous and reproducible validation of the introduced techniques with malware analysis experts.

This paper is organized as follows (see Fig. 2 for a graphical overview): Section 2 provides background information about the problem domain and provides a summary of our previous work on problem characterization and abstraction in relation to malicious software analysis. Section 3 presents related work in problem-oriented visualization research, malware analysis, and knowledge-assisted visualization. Section 4 provides a detailed description of our KAMAS prototype and Section 5 demonstrates the application of KAMAS in a usage scenario. Section 6 describes the used validation methods and the associated results. Finally, Section 7 is a reflection on the results of our research combined with the conclusion and future work.

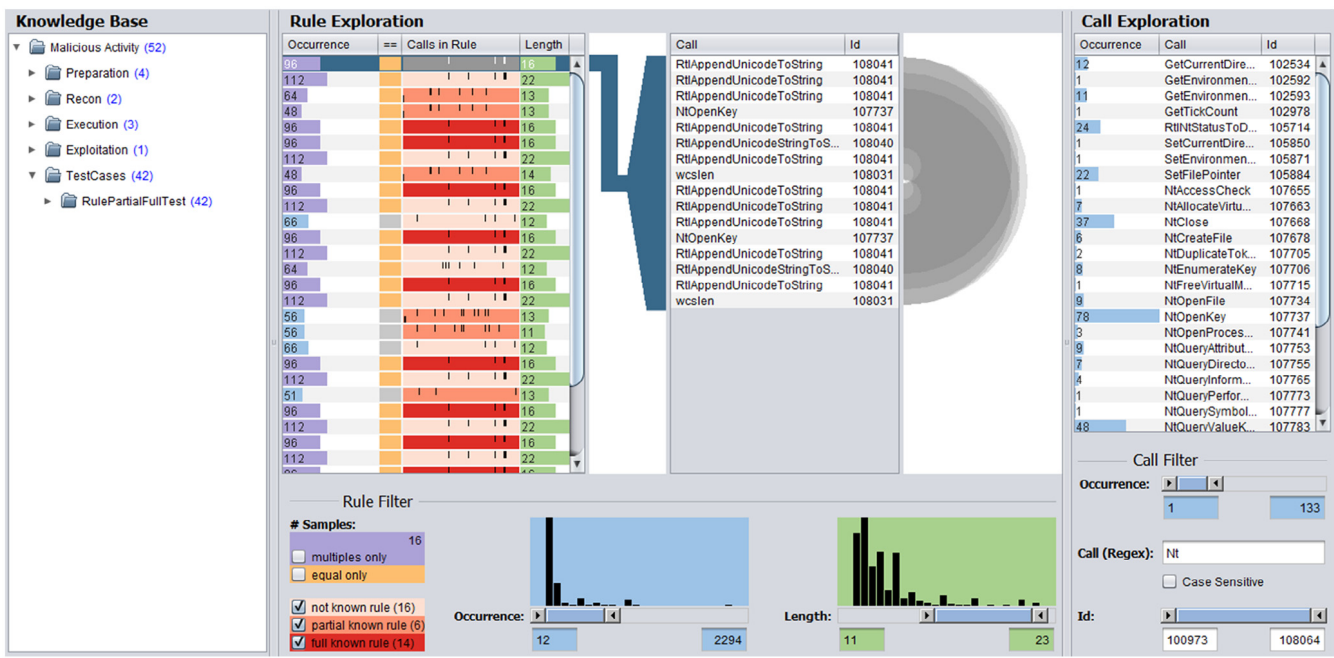

**Fig. 1 – Screenshot of the KAMAS interface during the exploration of a cluster grammar that includes combined system and API call sequences (rules) gathered during the execution of malicious software to find malicious subsequences (patterns).**

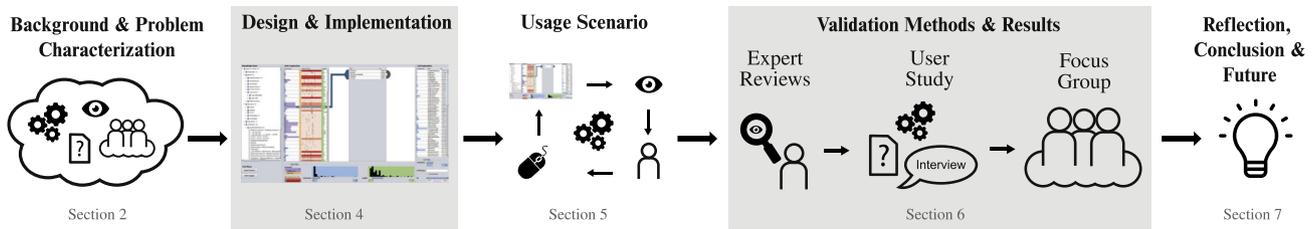

**Fig. 2** – Graphical overview of the KAMAS design study project reported in this paper.

## 2. Background and problem characterization

Throughout our design study project, we followed the well-known *nested model for visualization design and validation* as proposed by Munzner (2009). This systematic approach splits visualization design into four levels (*domain problem and data characterization*, *operation and data type abstraction*, *visual encoding and interaction design*, *algorithm design*) in combination with corresponding evaluation methods. Our problem-oriented approach to the study of knowledge-assisted visualization systems is based on our prior work (Wagner et al., 2014) which analyzed the needs of malware analysts in relation to their work on behavior-based malware pattern analysis. For this, we started with a problem characterization (Sedlmair et al., 2012) corresponding to the first level of the nested model, whereby we followed a three-step qualitative research approach including systematic literature research, focus group meetings and semi-structured interviews.

1. In the **literature research** (Section 3), it became apparent that visualization methods are less common in currently available malware analysis approaches.
2. During **focus group meetings** with IT-security specialists, we established a common understanding of their field of work and their objectives. In addition, a number of possibly applicable visualization techniques were discussed.
3. In the **semi-structured interviews** sessions, it was possible to identify preferred visual representation combinations (e.g., multiple views (Gotz et al., 2012) and arc-diagram (Wattenberg, 2002) or word tree (Wattenberg and Viegas, 2008)). Additionally, participants explained a number of currently available tools for the different work steps involved in behavior-based malware analysis.

Based on the gained insights, we analyzed the data, the users and the tasks in the problem domain using the data-users-tasks analysis framework introduced by Miksch and Aigner (Miksch and Aigner, 2014). This corresponds to the second level of the nested model (operation and data type abstraction).

### 2.1. Domain problem and data characterization

There are basically two different approaches for the detection of malicious software: the signature-based and the behavior-based approach (Dornhackl et al., 2014). Current malware detection/classification systems commonly use a signature-based approach which describes known malware by simple patterns (e.g., defined by regular expressions) or its syntactic characteristics (mostly bit strings) (Christodorescu et al., 2008). In general, the signature-based detection approach has several shortcomings (Christodorescu et al., 2008). On the one hand, obfuscation techniques commonly utilize polymorphic or metamorphic mutation to generate an ever-growing number of malware variants, which are different in appearance but functionally identical. On the other hand, signature-based techniques can only detect already identified and analyzed malware, otherwise they are overlooked. Since the signature-based approach can be used only for known malware, other techniques must be applied.

To overcome these shortcomings, the so-called behavior-based approach can be used. Here, a sample's activity is analyzed during execution using dynamic analysis techniques. These apply a previously defined set of rules to decide whether a sample's behavior is malicious. Behavioral analysis is a promising approach for detecting and pre-classifying malware: malware is not characterized by its syntactic appearance, but rather by its dynamic behavior.

**Behavior-based malware recognition:** Malicious behavior in software is identified through static or dynamic analysis (Egele et al., 2008). On the one hand, when statically analyzing a possibly malicious software sample, the binary file is usually disassembled and dissected function by function. Dynamic analysis, on the other hand, observes the sample's behavior directly (Egele et al., 2008): The supposed malware sample is executed inside an isolated, sometimes virtualized laboratory environment, which is instrumented with selected data providers, in order to generate a report (i.e. trace) of system and API calls sequentially invoked by the sample. These data providers are described in detail in Section 3 of a survey by Wagner et al. (2015). Both approaches (static and dynamic analysis) yield patterns and rules, which are later used for malicious software detection and classification.

**Pattern extraction process:** The behavior pattern extraction process (see Fig. 3) is usually split into several stages (Wagner et al., 2014) and can be based on different data providers (Wagner et al., 2015). Our collaborators cluster system and API call traces using Malheur (Rieck, 2016) concatenate traces of each cluster, and search patterns using the compression algorithm Sequitur (Nevill-Manning and Witten, 1997), which results in a context-free grammar. The patterns found are called "rules" by the collaborating domain experts, since the rules are parts of a context-free grammar. A rule describes a sequence of system or API calls with a length from

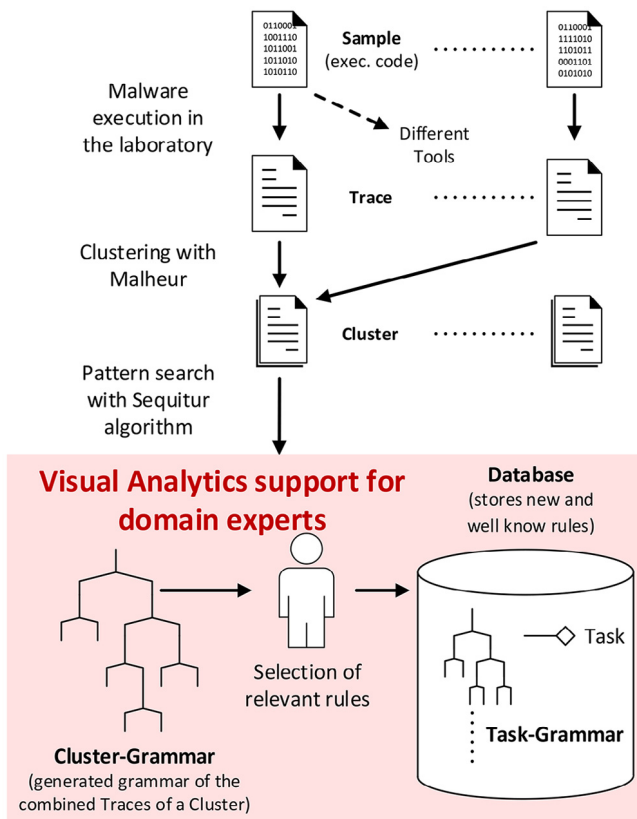

Fig. 3 – **Behavior-based malware analysis process as conducted by our collaborators.**

1 to *n* depending on the compression algorithm settings. The Sequitur algorithm is a lossless compression algorithm, i.e., the originally logged sequence of system and API calls can be reconstructed from the cluster grammar. See Section 4 for more details. A human analyst must then assess this grammar and extract rules describing potentially relevant behavior.

**Need for visualization:** Here is where visualization is needed to support the IT-security experts: The selection of relevant rules relies heavily on the accessibility and readability of the extracted information. Once a relevant rule has been identified by domain experts, it is stored in a pattern database. In the pattern database, a rule is assigned to a specific task of the hierarchical malicious behavior schema introduced by Dornhackl et al. (2014). This schema describes different types of malicious activities to enable the identification and the categorization of malicious software based on its high-level goals.

### 2.2. Terms and definitions

In this section we define terms related to the malware analysis process (Fig. 3) (Dornhackl et al., 2014; Wagner et al., 2014):

**Sample:** An executable program that is assumed to be malicious.

**System and API call:** The way a computer program requests a service (a function) from the operating systems kernel to be executed.

**Trace:** The behavioral execution logs of a sample in the correct execution order, e.g., a sequence of system and API calls.

**Cluster:** A set of traces (log-files) of *n* related samples in one file.

**Cluster-grammar:** The context-free grammar extracted by the Sequitur algorithm from the cluster file.

**Rule:** A grammar element, i.e. a sequence of system and API calls that occur in the generated cluster-grammar.

**Task-grammar:** An identified relevant rule, stored in the KDB assigned to a task of the malicious behavior schema (Dornhackl et al., 2014). The task-grammar is the foundation of the automatically generated rules, which are used to detect malicious behavior in newly loaded clusters.

**KDB:** The knowledge database storing the extracted implicit knowledge of the IT-security experts as explicit knowledge which is used for the automated analysis and visualization.

### 2.3. Operation and data type abstraction

Summarizing our results we could abstract the parse tree (the data structure which specifies the result of pattern search) of a cluster-grammar as a simple directed acyclic graph with nominal data attributed to the nodes. They consist of individual rules which in turn are composed of sequences of system and API calls. These have a cardinality typically greater than 100. The users are malware analysts (domain experts) whose main tasks are to select different rules, categorize them by their task and store them in the database as well as manual adaption and/or tuning of found rules (Wagner et al., 2014).

Based on these insights, we defined four key requirements (R) which have to be fulfilled by the KAMAS prototype:

**R1 Data:** *Handling of complex data structures in behavior-based malware analysis.* To ensure the effective analysis of a cluster-grammar, a large and complex data structure needs to be modeled, stored, and analyzed. This includes detailed information about the contained rules (Wagner et al., 2014) in the form of a directed acyclic graph with nominal attributes.

**R2 Visual representation:** *Visual representations appropriate for IT-security experts.* For the most parts, analysts investigate the collected data manually because the available tools do not cover all of their needs. In a preliminary study, we found that malware analysts preferred visualization concepts containing multiple views and arc-diagrams or word trees (Wagner et al., 2014).

**R3 Workflow:** *Workflow-specific interaction techniques.* The analysis workflow for behavior-based malware analysis contains several preprocessing and analysis steps (see Fig. 3). In relation to that, it is important to provide familiar interaction techniques to the experts, supporting them during the analysis in finding new malicious patterns and gaining new insights from the data.

**R4 Expert knowledge:** *Externalization of expert knowledge to reuse and share.* When analysts solve real world problems, they have large volumes of complex and heterogeneous data at their disposal. By externalizing and storing of the experts' implicit knowledge (Chen et al., 2009), it can be made system-internally available as computerized knowledge to support further analysis or other analysts.

We designed the visualization and interaction techniques of KAMAS, focusing on the defined requirements, followed by the algorithm design and implementation based on a user-centered design process (Sharp et al., 2007). In this way, we addressed the third and fourth level of Munzner's nested model (Munzner, 2009). During these steps, we collaborated with a group of three IT-security experts to get formative feedback for the design and implementation process of KAMAS. Two of them were malware analysts who were also involved in the previous focus group for problem characterization and abstraction (Wagner et al., 2014). All domain experts had more than 5 years experience in behavior-based malware analysis and experience with several tools. For summative evaluation of usability (Cooper et al., 2007), user studies were performed with predefined datasets provided by our IT-security collaborators.

## 3. Related work

In general, problem-oriented research is rather underrepresented in literature (Lam et al., 2012; McKenna et al., 2015, 2016; Pirker and Nusser, 2016; Sedlmair et al., 2012). One example is RelEx (Sedlmair et al., 2012), a design study with automotive engineers to optimize in-car communication networks. Other than that, Pretorius and Van Wijk (2009) point out that it is important for visualization designers to ask themselves: "What does the user want to see?" and "What do the data want to be?" as well as how these two points mutually enhance one another. In the problem domain of cyber security, Fink et al. (2009) developed a set of visualization design principles with a focus on high-resolution displays and presented prototypes according to these design principles. Goodall et al. (2004) conducted contextual interviews to gain a better understanding of the intrusion detection workflow and proposed a three-phased model in which tasks could be decoupled by necessary know-how to provide more flexibility for organizations in training new analysts. However, none of these user-centered studies tackled behavior-based malware pattern analysis.

Previous research explored the area of malware analysis from different points of view. Conti (2007) dedicated a part of his book to visualization for malware detection but focused on the network level. Lee et al. (2011) proposed a context-specific variant of scatter plots to classify malware by properties extracted using a signature-based approach. Dornhackl et al. (2014) introduced a workflow for malware pattern extraction. They log the system calls of malicious software which is executed on a host and analyze them. In this way, it is possible to extract and define malicious system call patterns. In 2015, Wagner et al. (2015) presented a survey on visualization systems for malware analysis. They described 25 malware visualization tools and categorized their main analysis focus based on the "Malware Visualization Taxonomy". This taxonomy described the two main areas of malware visualization systems. On the one hand, there are systems for "Individual Malware Analysis" which support the analysis of a single malware sample to gain new insights (e.g., Donahue et al., 2013; Wüchner et al., 2014). On the other hand, there are systems for "Malware Classification" which support the comparison of many samples to identify the common behavior (e.g., Gove et al., 2014; Han et al., 2014; Long et al., 2014). For example, Wüchner et al. (2014) proposed the interactive "DAVAST" system for the detection and analysis of email worm attacks. Additionally, Gove et al. (2014) introduced "SEEM", which allows the behavioral comparison of a large set of malware samples in relation to the imported DLLs and callback domains.

However, none of these approaches covered the full set of requirements as specified by Wagner et al. (2014) during their problem characterization and abstraction. Our paper is based on this prior research, with the goal of developing an analysis interface which reflects the analysts' workflow. In particular, we integrated a set of well-known visualization techniques for the IT-security community and integrated a KDB structured by the "malicious behavior scheme" (Dornhackl et al., 2014). Thus, KAMAS will contribute to the emerging topic of knowledge-assisted visualization (Chen et al., 2009). Some problem-oriented projects (e.g., Mistelbauer et al., 2012) and general frameworks (e.g., Tominski, 2011) integrate explicit knowledge in visualization but not in a form similar to malware behavior patterns.

## 4. Design and implementation

This work is based on our prior work (Wagner et al., 2014) to characterize and abstract the domain problem of behavior-based malware analysis. Therefore, we set up a design study project to find a visualization solution that followed a user-centered design process (Sharp et al., 2007). We involved a group of three domain experts in malware analysis and IT-security to keep the design in line with the analyzed needs of our prior work. We iteratively produced sketches, screen prototypes, and functional prototypes (Kulyk et al., 2007). Based on these, we gathered feedback about the design's usability and how well it supports their analysis needs. The design study resulted in the KAMAS prototype (see Fig. 4), which is implemented in Java. Next, we elaborate on central design decisions.

**Visualization concept:** When loading an input file, all of the included system and API calls are presented in the "Call Explorer" area (see Fig. 4.3).

*Call explorer:* The call table (see Fig. 4.3.a) contains three columns. From left to right, it provides at first detailed information on the call occurrence, which shows the appearance in the loaded file, thus helping the analyst to identify interesting calls depending on their occurrence frequency (e.g., if the cluster was generated with 16 samples, calls that occur 16 times or a multiple of 16 times are more interesting than others). The occurrence frequencies are quantitative data presented as bar charts, which serve as a quick visual decision support, and numbers for a precise comparison (graphical tables). The second column visualizes the name of the calls as a string which is a nominal value and the third column displays the unique ID of the call as an ordinal value.
*Rule explorer:* The included rules of the loaded input file are visualized in the "Rule Exploration" area (see Fig. 4.2) found in the rule overview table (see Fig. 4.2.a). From left to right, "Occurrence" tells the analyst how often the visualized rule occurs in the loaded file visualized by bar charts and numbers. These values are related to the occurrence histogram by color (see Fig. 4.2.g) in the "Rule Filter" area, which serves as an overview of the rule distribution depending on their occurrence. The second column "⚌", provides information as to whether the rule is equally distributed in all the traces of the loaded cluster. More precisely, if the represented rule occurs equally often in all traces combined in the cluster-grammar the related column will be colored orange, otherwise it is colored gray. In the third column ("Calls in Rule") the graphical summary (see Fig. 5) visualizes the included calls.

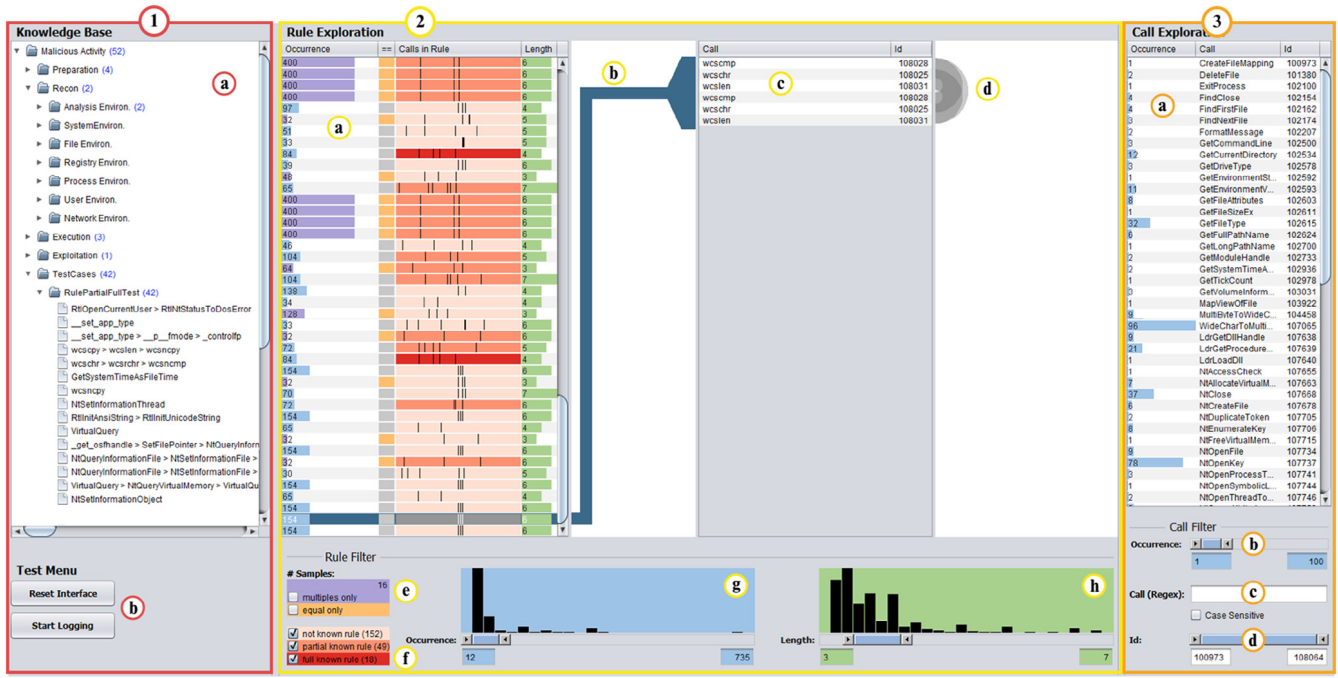

**Fig. 4 – User interface of the KAMAS prototype with its three main areas. 1) The (1.a) tree structure of the knowledge database (KDB) and the (1.b) test menu used for logging during the user study. 2) The rule explorer including the (2.a) rule overview table and the graphical summaries, the (2.b) connection line to the (2.c) rule detail table representing the included calls and the (2.d) arc-diagram for pattern recognition support. Additionally, on the bottom there are different filtering options (2.e–h). 3) The call explorer interface including (3.a) the table of calls available in the loaded file and different filtering options (3.b–d).**

*Graphical summary in the rule explorer:* Due to the fact that a rule consists of 1 to *n* calls, it is difficult for the analysts to read them completely, to remember them in their correct order and to compare them with other rules. To simplify the unfavorable textual representation and to make the structures of the rules more comparable, we developed a graphical summary for the representation of the contained calls. In the graphical summary, each pixel is related to a specific system call ID and has a fixed position to allow their comparison. Each call ID related pixel signifies whether a call of a certain type is present in the rule (colored black) or not. Thus, fingerprints of rules that have many calls in common look visually similar. If a rule's call occurs more often, only one line is displayed for it. For example, if a rule contains four calls (length of four) and less than four lines are displayed in the graphical summary (e.g., three or two), one or more calls occur more often than once. Based on the graphical summary, the analyst has the opportunity to compare the included rules and to recognize similar rules faster in relation to the included calls. Furthermore, the graphical summary is based on the Gestalt principle of similarity (Johnson, 2014) in order to support the analyst in quickly recognizing related call combinations. That is to say, rules that contain similar sets of calls have similar line patterns. This allows analysts to quickly assess similarity without the need to consider exactly which calls are part of the involved rules. Additionally, a tool tip is provided for a quick look at the included calls in the rule. The three different color intensities from dark to light red of the graphical summary informs the analyst whether a rule in the KDB is "fully known" (already stored), "partially known" (partially stored) or "not known" (unknown) whereas these rules are defined by the analyst or other users. A "fully known" rule which is highlighted dark red, is included as-is in the KDB. A "partially known" rule (middle red) describes a rule which contains at least one more call at the beginning or at the end of a "fully known" rule, and is not stored in this form in the KDB. The highlighting of the rules is performed internally by comparison of the rules contained in the input file and the rules stored in the KDB. In the last column, the analyst receives information about the length of the visualized rule (number of included calls). If the analyst compares the number of lines from the graphical summary with the related rule length, the analyst can recognize quickly whether it matches or not as described above. Like the occurrence column, this column is also related to a histogram (see Fig. 4.2.h) in the filter area showing the distribution of the lengths of the rules.

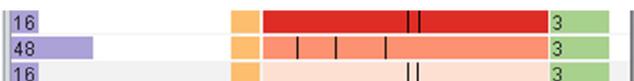

**Fig. 5 – Detail of the graphical summary in the rule overview table. This is a space-efficient rule representation with lines marking distinct calls contained in the rule, regardless of the occurrence order and number.**

*Rule detail table in the rule explorer:* Additionally, in the "Rule Exploration" area, the rule detail table (see Fig. 4.2.c) visualizes the included calls and the unique ID of a selected rule in its actual execution order. To support the analysts during their work on pattern search, we included a real time pattern search. For the visualization of patterns which are included in the represented execution order, arc-diagrams (Wattenberg, 2002) are used (see Fig. 4.2.d). In this way, the analyst receives a visual representation of recurrence patterns up to the five largest patterns in a rule.

*Connection line in the rule explorer:* To connect the rule overview table and the rule detail table, we integrated a table connection line (see Fig. 4.2.b). This line connects a selected rule from the overview table to the detailed execution order of the integrated calls in the detail table. Thus, the user is informed of where the selected rule is located during scrolling in the view. By a click on the connection line, the selected rule comes back into focus.

**Visual interface design concept:** IT-security experts are well skilled programmers and programming IDEs usually have a similar structure and workflow. Based on the findings of our prior research described in Section 2, we decided to imbue the design of KAMAS with programming IDE interfaces such as Eclipse or NetBeans. Based on this interface structure we could establish a familiar workflow concept on multiple views for the IT-security experts. In relation to these well-known interface structures, we situated the tree structure of the KDB (see Fig. 4.1) to the left side of the interface like the project structure. Additionally, we positioned the "Rule Explorer" in the center of the screen (see Fig. 4.2) like the development area. This is the most used element of the prototype – the main screen, used to explore and analyze the calls which are included in a rule. Moreover, we positioned the "Call Exploration" area to the right of the interface (see Fig. 4.3), like the functions overview area in commonly used programming IDEs. In this area, the user has the ability to explore all of the calls which are included in the rules in the middle and to obtain detailed information about them. In addition to the design decision in relation to a programming IDE, we used Gestalt principles (Johnson, 2014) to improve interface clarity. Each exploration area (Rule Explorer and Call Explorer) contains its own filtering area below the data visualization (based on the Gestalt principles of proximity and similarity).

**Interaction concept:** For a better understanding of its functionality, we describe KAMAS according to five steps based on the visual information seeking mantra by Shneiderman (1992): overview first, rearrange and filter, details-on-demand, then extract and analyze further.

*Overview:* When the malware analyst loads an input file, the tables for rule overview (see Fig. 4.2.a) and call exploration (see Fig. 4.3.a) are filled with the complete data of all extracted rules and all calls occurring in these rules. Furthermore, the histograms (see Fig. 4.2.g and Fig. 4.2.h) give an impression of the distribution in rule occurrence and length of the currently loaded cluster grammar file.

*Rearrange:* The analyst can now rearrange both the display of rules and calls by sorting the tables by any column.

*Filter:* The next step is to reduce the number of rules under consideration. For this purpose, the interface offers a selection of several filtering options. The analyst can start by filtering the calls (see Fig. 4.3.b–c) or selecting calls directly from the call exploration table. Furthermore, the analyst can filter rules by their number of occurrence, by their length, whether their occurrence is equally distributed across samples, and whether they match or partially match rules from the KDB (see Fig. 4.2.e–h). The rules displayed in the center table are updated immediately, where the graphical summary gives an impression of the included calls.

*Details-on-demand:* If a rule catches the analyst's interest, it can be selected from the rule overview table. This action opens the rule in the rule detail table (see Fig. 4.2.c), where the analyst can read the included calls in their sequential order. The arc-diagram provides information about repeated subsequences within a rule (see Fig. 4.2.d). All the contained subsequences are analyzed in real time. In order to not confuse the analyst, only the five largest subsequences are presented in the visualization.

*Extract:* Once the analyst has found a possibly malicious rule, it can be added to the KDB by dragging it from the rule overview table into the tree structure of the KDB (see Fig. 4.1.a). Alternatively, the analyst can select some calls of a rule in the rule detail table and add them to the KDB by drag and drop. This updates the background color of the rule overview table, which allows further analysis under consideration of the added rule.

**Input data:** In general, the input data are sequences of system and API calls which are logged during the execution of malware samples in protected laboratory environments (Wagner et al., 2015) as described in Section 2. In some preprocessing steps, the data of several analyzed malware samples (e.g., from the same malware family) are clustered and transformed into a context-free grammar in order to reduce the immense number of calls, simplify the analysis and reduce storage costs. The outcome of these preprocessing steps is loaded into KAMAS.

In our specific case, the following preprocessing steps are adopted from our collaborators' malware analysis process (see Fig. 3): First, the sample under scrutiny is executed inside isolated, partially virtualized laboratory environment using a number of different tools (e.g., APImon (API Monitor, 2016), Procmon (Microsoft, 2016)). By monitoring all activities, these systems generate a report (i.e. trace) of system and API calls sequentially invoked by the sample. In the second step, the traces are clustered with Malheur, an automated behavior analysis and classification tool developed by Rieck et al. (Rieck, 2016). In the third step, all traces within a cluster are concatenated and processed using the Sequitur algorithm (Nevill-Manning and Witten, 1997). In relation to this step, the system and API call parameter are neglected (e.g., memory addresses, user names) which simplify the creation of a context-free grammar of the combined samples in the cluster and generates a bird's-eye view. If some parameters are needed, they can be considered by attaching them behind the call using a separator (e.g., CreateProcess#winword.exe). It is important to note that the alphabet of the context-free grammar will increase by considering the call parameters. Originally developed for file compression, Sequitur automatically replaces frequent patterns with short symbols, effectively generating a context-free grammar in the process, which is referred to as cluster-grammar and forms the input data format for the KAMAS prototype. The data structure of the cluster-grammar is a simple directed acyclic graph. Leafs are called "terminals" and represent distinct system calls. Other nodes are called "non-terminals" and are represented by numbers. Additionally, a sequence of system calls (terminals) composed from the cluster-grammar is called "rule" and represents a part of the sample's actual execution order.

While our collaborators only analyzed malware samples on Windows operating systems, it will also be possible to analyze malware for other operating systems based on data providers for these operating systems. After the transformation into the input data structure as described above, they can be loaded into KAMAS.

**Internal data handling concept:** To increase the performance of our prototype, we decided to use a data-oriented design (Fabian, 2013) (e.g., used in game development and real time rendering) to organize and perform transformations on the loaded data. For example, we implemented map and set data structures as translation tables and for the input data, we mapped the strings to integers. Based on this mapping, we built an internal data structure to increase performance and to decrease memory usage. In combination with these structures, we implemented an action pipeline (similar to a render pipeline) in which it is possible to link filter options in order to realize high performance dynamic query environments and to perform all operations on the data in real time.

To test the robustness and performance of KAMAS, we loaded and worked with several different cluster-grammar files containing between ten and 42 traces of malware samples in the cluster-grammar at sizes between 61 and 7278 rules. The included malware samples were collected by our collaborators from the IT-security department in 2014 to test their behavior-based malware analysis and clustering system. Some analysis examples and KAMAS prototypes are available on: http://phaidra.fhstp.ac.at/o:1264. Overall, they collected a sample set with 800 different malware samples of different families (worms, trojanhorses and bots). Additionally, with these datasets, the system handles more than 8000 different windows system and API calls.

**Filter possibilities:** The implemented filtering options are organized in two separated filter action pipelines where the input data depends on the loaded analysis file. The first pipeline connects all filters for the call exploration area by using an "and" operator. The result of the first filter pipeline affects the call exploration (see Fig. 4.3) and the rule exploration (see Fig. 4.2) area. The output of the call exploration action pipeline is the basic input of the rule exploration action pipeline. In this pipeline, all included filters work similar to the call ex-ploration action pipeline and control only the rule exploration area of the interface (see Fig. 6). The integration of action pipelines for data filtering has the benefit of being easy to change, and quickly include or exclude new or not needed filtering options into KAMAS.

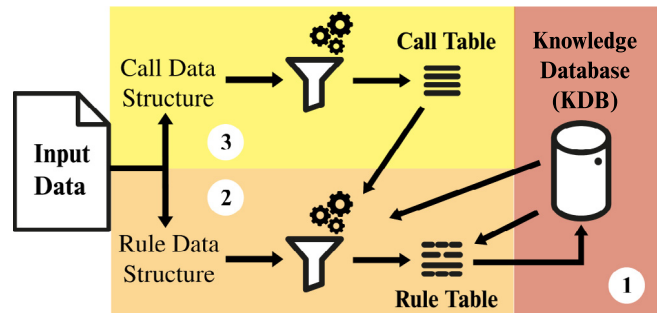

Fig. 6 – Data flow diagram of the KAMAS prototype including the two filter action pipelines. 1) The KDB which affect the rule table depending on the stored knowledge. 2) The rule table representing the included rules of the loaded input data. Its filter pipeline gets affected by the output of the KDB and the call table filters. 3) The call table visualizes the included calls. Its filtering pipeline affects the call visualization and biased the rule table filter pipeline.

*Call filters:* Below the call table, three filter options are arranged. All these filters determine the contents of the call table and subsequently the rule exploration. The "Occurrence" filter (see Fig. 4.3.b) is implemented as range slider to select the region of interest. Below is the "Call (Regex)" filter (see Fig. 4.3.c) with a separate option for "Case Sensitive" search. This filter allows the entry of a regular expression or plain strings which are internally converted into a regular expression for a sub-string search. The third filter option operates on the unique 'ID's (see Fig. 4.3.d) of the calls. If the user selects one or more calls in the call table (see Fig. 4.3.a), a fourth filter level will be established. All of these filters are connected in a dynamic query system (see Fig. 6.2).

*Rule filters:* From left to right, we included the following filtering options according to the suggestions of the collaborating domain experts: The "multiples only" option provides the opportunity to show only rules that occur with a multiple number of times of the included samples in the loaded cluster (e.g., 16 samples included → only rules that occur 16, 32, … times will be shown). The option "equal only" eliminates all rules which are not equally distributed in the included samples (see Fig. 4.2.e). Below the knowledge database filter options are arranged (see Fig. 4.2.f) to activate or deactivate the visualization of rules based on their knowledge state. Additionally, the rule exploration filtering unit includes two range sliders (Ahlberg and Shneiderman, 1994) combined with histograms (see Fig. 4.2.g–h) for the visualization of the data distribution and for the selection of a region of interest. Fig. 4.2.g serves the filtering option based on the "Occurrence" of the rules and the related histogram shows the distribution of the rules in relation to the loaded file. The user can also select a region of interest depending on the "Length" (see Fig. 4.2.h) by range slider interaction, similar to the "Occurrence" workflow.

**Externalized knowledge integration:** To support the malware analysts during their work, we integrated a KDB related to the malware behavior schema by Dornhackl et al. (2014)., which is included on the left side of the interface (see Fig. 4.1) as an indented list (tree structure). At the end of each folder description, the number of contained rules in the integrated subfolders is shown. To add new rules to the KDB, the system provides two possibilities: On the one hand, the user can add a full rule by drag and drop from the rule overview table (see Fig. 4.2.a), and on the other hand, the user has the ability to select a set of calls from the rule detail table (see Fig. 4.2.c) and add them by drag and drop to the KDB. When adding a new rule, the tree automatically unfolds the hovered folder. All of the rules which are integrated in the KDB will be checked against the loaded input data automatically. In this way, the system distinguishes between three types of rules (see Fig. 4.2.a and 4.2.f): not known (light red), partially known (middle red) and fully known rules (dark red). Additionally, the KAMAS prototype provides the ability to activate or deactivate (parts of) the KDB or to select the types of rules which should be shown in the rule overview table.

*Knowledge generation loop:* Fig. 7 provides an overview of the system's knowledge generation loop, starting at the dark gray inner loop. In general, the system's KDB stores all known rules, which were generated by former cluster file analysis sessions. If the analyst loads a new cluster file, the included rules will be checked automatically against the activated KDB parts (see Fig. 7.1). Depending on the prior automated analysis, the system provides a visual representation of loaded rules in relation to its knowledge state (see Fig. 7.2). Henceforth, the analyst can carry out the data exploration and analysis (the analyst is part of the knowledge generation loop). During the cluster analysis, the analyst has the ability to extend the KDB with new rules (see Fig. 7.3) found during the data exploration process. By adding new rules to the KDB, the system automatically refreshes the rules highlighting depending on the new knowledge state (see Fig. 7.4), which brings us into the outer (light gray) loop. Here the analyst is part of the continuously recurring loop (see Fig. 7.5) (for data exploration see Fig. 7.6 and for knowledge generation see Fig. 7.7).

## 5. Functional overview of KAMAS

If an analysts loads new malware data, KAMAS provides a general overview of all included system and API calls and all preprocessed rules which are sequences of calls (see Fig. 4). Furthermore, in the "Rule Explorer" (see Fig. 4.2.a) the analyst can see a graphical summary with different background colors depending on the knowledge state of the KDB. If a rule is fully known in the KDB, the rule's background is dark red, a light red background tells the analyst that a KDB rule is included in the represented rule but surrounded by at least one other call. The light red background tells the analyst that this rule did not match with the KDB, which may mean that it is not hostile or was not yet recognized. Based on the different filtering options provided, the analyst has the possibility to search for rules of interest iteratively.

**Rule selection and investigation:** If the analyst selects a rule in the "Rule Explorer", the selected rule becomes highlighted (see Fig. 4.2.a). At the same time, on the right side, a detail table is generated which represents the individual calls in chronological sequential order (see Fig. 4.2.c). Additionally, a vertical arc-diagram (see Fig. 4.2.d) represents sub-sequence repetitions found, which supports the analyst in finding the smallest sequence without sub-patterns.

**Searching for specific calls:** If the analyst searches for a specific call, the call's name can be entered into the name filter in the "Call Filter" area (see Fig. 4.3.c). Here, the analyst has the ability to enter regular expressions for the search. Thus, it is also possible to add only a part of the call's name to search for a specific group. Additionally, it is also possible to select one or more calls of interest in the "Call Exploration" area (see Fig. 4.3.a). All of these abilities affect the "Rule Exploration" area so that each represented rule has to contain at least one of the selected (filtered) calls.

**Storing new patterns:** If the analyst wants to store a new rule in the KDB, the system provides two options. The first option is that the analyst selects the full rule in the overview table and adds this rule by drag and drop into the KDB (see Fig. 4.1.a). The second option relates to the detail table. The analyst selects only the calls of interest of a found rule and adds them as a new rule by drag and drop to the KDB. No matter which option the analyst selects, the tree structure in the KDB automatically unfolds the hovered categories so that the analyst can add the selection in the right category or sub category.

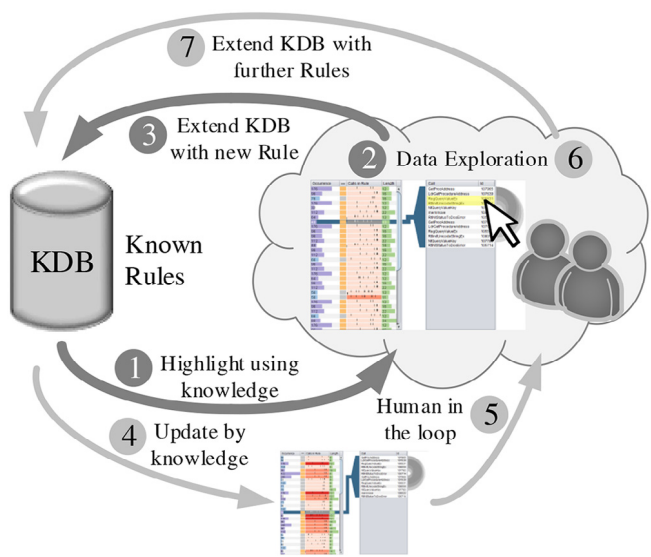

**Fig. 7 – Overview of the system's knowledge generation loop, beginning with the dark gray inner loop after loading a new file and changing to the light gray outer loop for interactive data exploration and knowledge generation cycle. The analyst is a major part in both loops.**

## 6. Validation methods and results

To validate the KAMAS prototype and provide evidence for its effectiveness, we followed a threefold research approach consisting of moderated expert reviews, user studies and focus group meetings (Lazar et al., 2010). All of the insights were documented in detail to ensure reproducibility (Smuc et al., 2015) and used to improve our research prototype. All materials used, such as interview guidelines and tasks, are available on our supplement material homepage: http://phaidra.fhstp.ac.at/o:1264.

### 6.1. Expert review

In the first step, we conducted moderated expert reviews to eliminate usability issues in the basic functionality and appearance of the interface.

**Participants:** To validate the visual interface design, we invited three usability experts for this step (see Table 1). Each of them has between two and 20 years of experience in this field. Two of them are between 20 and 29 years of age, have a Master's degree and advanced knowledge in usability. One of them is between 40 and 49 years old, has a PhD degree and expert knowledge in usability engineering from industry and research projects.

**Design and procedure:** Each usability expert received a short introduction to the basic features and the workflow of the system. Next, each expert was walked through each feature individually and was asked to critique potential usability issues.

**Apparatus and materials:** As evaluation material, we generated a fully functional build of KAMAS and used the same version for each expert review. The review sessions were performed on a 15″ notebook with a full HD screen resolution and an external mouse for navigation. Each expert review was documented on paper by the facilitator.

**Results:** The basic color scheme of the prototype was found to be easily recognizable. Only the coloring of the "multiples only" option was pointed out as being not well differentiated from the other elements (see Fig. 4). Basically, the dynamic query features (Ahlberg et al., 1992) were described as being very useful. If the user left the focus of a filtering input box, the interface had to automatically apply the entered parameter. Additionally, in the filtering option "call", the experts suggested that it would be helpful to update the search results after each input. Overall, all of the usability experts provided positive feedback on the design structure of the prototype. All of the expert's suggestions were included for a redesign and revision of the prototype in order to prevent the users from having basic interface issues.

**Table 1 – Overview of usability experts who participated in the expert reviews of the KAMAS prototype. (Knowledge: 1: = basic, 2: = skilled, 3: = advanced, 4: = expert).**

| Person | Age | Knowledge | Gender | Education |
|---|---|---|---|---|
| E1 | 40–49 | 4 | m | PhD |
| E2 | 20–29 | 3 | f | MSc |
| E3 | 20–29 | 3 | m | MSc |

**Table 2 – Data of the user study participants. (R: = research group, F: = faculty; knowledge: 1: = basic, 2: = skilled, 3: = advanced, 4: = expert).**

| Person | Organization | Age | Knowledge | Education |
|---|---|---|---|---|
| P1 | R | 30–39 | 4 | PhD |
| P2 | R | 30–39 | 4 | MSc |
| P3 | R | 30–39 | 4 | MSc |
| P4 | R | 30–39 | 4 | PhD |
| P5 | R | 30–39 | 3 | MSc |
| P6 | F | 60–69 | 4 | PhD |

### 6.2. User study

A user study with six IT-security experts was performed in October 2015. It lasted one hour on average and encompassed five analysis tasks, the system usability scale questionnaire (SUS) (Brooke, 1996), and a semi-structured interview. The goals (G) and non-goals (NG) of the user study were defined as:

**G1:** Testing the functionality of the research prototype.
**G2:** Testing the visualization techniques for comprehensibility in relation to the domain.
**G3:** Testing the utility of knowledge storage and representation in the system.
**NG1:** Comparison of the research prototype with another analysis system.
**NG2:** Conducting performance tests.

**Participants:** We invited six IT-security experts (see Table 2) to participate in the user study. Two participants were also members of the focus group for the user-centered design process and gave feedback on sketches and early prototypes (see Section 4). All subjects are working in the field of behavior-based malware detection and analysis or in a closely related field of IT-security. All of them are involved in malware analysis projects in cooperation with different industries partners.

**Design and procedure:** At the beginning of the user study, each participant was asked about a general impression of the user interface and which functions could be recognized. This step took approximately five minutes. Subsequently, each participant had to solve five guided analysis tasks. The first three included a step-wise introduction to the system and the last two were combined analysis tasks to determine the understanding of the prototype's workflow (this step was also required for the subsequent SUS questionnaire). Each analysis task was read to the participant at the beginning of the task, and for reference, each task handed over to the participant in printed form. For the analysis tasks, participants spent approximately 30 minutes.

After the analysis task session, each participant had to fill out a SUS questionnaire in less than five minutes. The SUS is a standardized, technology-independent questionnaire to evaluate the usability of a system (Brooke, 1996). During this test, the participants were asked ten questions on a five-level Likert scale from strongly agree to strongly disagree.

Finally, we performed semi-structured interview sessions with an average duration of 20 minutes. For this, we used an interview guideline consisting of ten major questions addressing general system usability, filtering, using the KDB, and individual visual metaphors used in KAMAS.

**Apparatus and materials:** We used a silent and clean room without any distractions to perform the three parts of the user study for all participants under the same conditions.

The five analysis tasks were performed on a 15″ notebook with full HD screen resolution and an external mouse. As dataset for the analysis tasks, we used a set of 16 malware samples (from the same family) executed and logged on a Windows operating system and transformed into KAMAS' input data format containing 794 rules as described in Section 4. To achieve the best outcome, we asked the participants to apply thinking aloud (Nielsen, 1993). We recorded the screen and the participant using the notebook's internal webcam. In parallel, the prototype logged user interactions using EvalBench (Aigner et al., 2013) (see Fig. 4.1.b) and the facilitator took notes in our pre-defined test guideline.

The SUS questionnaire and semi-structured interview were conducted on paper in the participants' native language. For the detailed questions, we used small images in the interview guidelines to support the participants in recalling the respective parts of the prototype. All test tasks used and interview guidelines are available on our supplement material homepage: http://phaidra.fhstp.ac.at/o:1264.

**Results:** The following part describes the major findings of each part of the test and summarizes their results.

*Analysis tasks:* During the five analysis tasks, all participants described their solutions to the tasks at hand by thinking aloud. They expressed problems as well as benefits of the systems during their work. A major problem pointed out by most participants was that in many cases they had to read tool tips about different filter options but the text disappeared too quickly [P1, P2, P3, P5]. Additionally, all participants told us that they had problems understanding the arc-diagrams. Only P3 stated: "It is an indication that something is included. I understood and applied it." In contrast, all but one participant noted that they quickly understood the handling of the knowledge database and its features in the prototype [P1, P2, P3, P5, P6].

*System usability scale (SUS):* By applying the SUS, we were able to obtain a good indication concerning the handling of our KAMAS prototype. The results show a SUS value of 75.83 points out of 100, which can be interpreted as good without significant usability issues according to the SUS description by Bangor et al. (2009). They described the SUS questionnaires result from different perspectives: From the perspective of the acceptability range, a value between 70 and 100 points is labeled as "acceptable and from the adjective rating perspective, the KAMAS SUS result lies in a range between good (73 points) and excellent (85 points). In general, all participants were able to recognize the interaction design and interface changes in relation to the work steps they fulfilled. Based on the SUS description and average evaluation of the system by the participants, the result of the usability assessment was very positive. Sauro (2011) compared 500 SUS scores and identified the average score as 68 points. Additionally, he showed that only 36% of the compared tests reached an SUS score higher than 74, and only 10% of the systems reached an SUS score greater than 80, which shows us that our system receives a grade of 'B' at this implementation state.

During the test, the participants again addressed the tool tip complication and the missing brief introduction of the included visualization techniques. These comments were also apparent concerning SUS question ten: "I needed to learn a lot of things before I could get going with this system" (Brooke, 1996), with a score of 50%. In contrast, the SUS questions concerning the complexity, functionality integration, consistency, learnability and usability (number two and five to eight) clearly reached more than 80%. Thus, focus areas for further improvements were identified.

*Interviews:* Here, we present the results of the performed interviews structured along ten main questions. All participant statements quoted in this section have been translated from German to English by the authors.

*Are the filter options understandable?* In general, the filter options were classified as very useful by all participants (P1–P6), e.g., as P6 stated, "Yes, nothing more to say about it." P4 and P5 also added that all filters were well labeled so that it was possible to recognize their mode of action immediately. In the call filter by name part, it was confirmed by the participants that the regular expression support was a great advantage. Additionally, P1 was very glad that the system automatically performs a full text search if there are no regular expression elements added. P1 and P3 added that the range sliders were very helpful and reflected the selected area very well. Likewise, it was mentioned that the arrangement is intuitive.

*Were the different visualization techniques helpful?* P1 and P3–P6 indicated that the various visual elements contributed significantly to understanding the system. P1 and P6 stated that the graphical summary (see Fig. 5) was very useful, although participant P6 added that a brief description would have been helpful. P3 mentioned that the interesting rules were immediately found by adhering to the graphical representations (e.g., coloring, bars in the table columns, arc-diagrams for pattern recognition) and they provided a good overview of how often they occur. P1 and P3–P6 noted that the colored background of the individual categories and elements aided significantly in finding relationships and helped in understanding the system. P1 and P6 indicated that the arc-diagram was not really understandable, and only P3 immediately recognized that it is used to visualize recurring patterns in the data. P3 added that the histograms were very helpful to see the loaded file's data distribution.

*Did you understand the use of the knowledge database?* The handling and the meaning of the KDB was assessed as very useful by all subjects (P1 – P6). However, P5 indicated that a short description of the functional advantage would have been helpful in understanding it more quickly. Likewise, it was described as very easy to insert new rules in the database. The tree structure was very well received by the participants. P3 stated that drag and drop worked very well for the insertion of new rules. Interestingly, P2 stated that "The knowledge database can also be interpreted as a kind of pre-filter for data use". This relates to the automated analysis by using the KDB to highlight the knowledge state of included rules in the loaded input file. Thereby, the analyst gets the ability to include or exclude these rules depending on the explicit knowledge.

*Were the different ways of knowledge representation helpful in finding the right choices?* In general, all participants (P1 – P6) stated that the KDB is ([P4] "probably") helpful for decision making. P1 and P5 described it as well suited for checking the rules in the loaded file. That answers frequently asked analysis tasks such as: "Is the sample doing what the database says?" [P1]; "What is the type of the considered sample?" [P1, P5]. P3 and P5 mentioned that the database would also be used as a search option to instantly find or investigate rules with predefined content. In this regard, P5 added that the KDB was very helpful for quickly searching and filtering. "If the Knowledge database is filled, it makes sense to focus on this" [P1]. P2 indicated that it depends on the task.

*How was the expert knowledge represented in the system?* P1, P3 and P4 related to the KDB and their tree structure representing the explicit knowledge stored in the system. Additionally, P2, P5 and P6 referred to the color gradation in the graphical summary representing the knowledge state of a rule in the KDB. It was very well understood by the participants, that the KDB could be filled by other experts or by themselves. Thus they had the chance to explore existing externalized knowledge, learn from it and add their own. "Any expert has added this into the database marked as important or prominent", stated P3. P2, P5 and P6 referred to the red three-step color map which is used for the background in the graphical summary. In this way, the user recognizes immediately if a rule matches with the database (not known, partially known, or fully known). The red three-step color map was understood by all participants and described as clear and logical. P1 added, "If a malware changes over time, it can be found by means of expert knowledge".

*Were the bar charts included in the table cells helpful?* The bar charts in the table cells (see Fig. 5) were generally described as very helpful and useful, because one immediately gets an overview of the distribution and thus can estimate the frequencies at a glance. Additionally, P1, P2, and P3 added that the sorting of the data can be recognized immediately, thus min and max values could be found very quickly. P1 – P3, P5 – P6 mentioned that the bar charts were very easy to read, helped recognizing frequencies, and in comparing the occurrence, length, and distribution. P4 and P5 added that this presentation is much better for comparison than numerical representations because they can be estimated. In case it is necessary to get the exact value of a cell entry, all participants stated that they would use the exact number.

*Were the histograms displaying rules helpful?* P1, P3, P4, and P6 stated that the histogram information on the distribution of the length and occurrence of the rules in relation to the loaded file was important additional information. P1 additionally mentioned that it would be helpful to gray-out the areas of the histogram which are not selected in relation to the underlying range slider. "A solution by means of shading would also be feasible and would require less space", stated P4. P5 indicated that such a representation is not applied in other known tools.

*Was the connection between the two tables helpful?* The connecting line was described by all participants as very or absolutely useful. P3 and P6 called the connection optically supportive because it helped the eyes' transition to the second table. Additionally, P1 and P6 valued the connection line as helpful for scrolling because you can always find the associated rule again. "After recently trying it out, the connection line was very helpful", stated P5.

*Were the arc-diagrams next to the detail table helpful?* The arc-diagram provoked mixed opinions among the participants. All of the participants stated that interesting elements in the resolved rule were made visible by the arcs. However, the precise usage and meaning, was recognized only by P3. "This is a part of a pattern in a rule which is again discoverable on the first glance. This might be confusing, but at least you always see if patterns are contained", stated P3. In general, four out of six participants (P1, P2, P4, and P5) considered this visualization method to be confusing and would likely reject it. They demanded a different form of presentation for pattern highlighting. Other critical comments were: "The overlap of same colors does not make sense" [P4]. "With a description certainly useful" stated P2 and P5. "I did not understand it without [visualization] background knowledge" [P5].

*What is your overall impression of the software?* In general, all participants (P1 – P6) described the prototype as very clear after having a brief explanation. The overall layout was described as clear, and the graphical summary was also found to be very helpful. "Simple and very good. Definitely useful for the application" stated P6. Likewise, it was stressed that the prototype includes many functions but it is not overloaded. P2 commented the system is useful for exploratory tasks in malware analysis. P5 added that a long-term test would be interesting. Participant P4 suggested changing the arrangement of the KDB and the call exploration (see Fig. 4) in a way that a workflow results from left to right.

**Combined results:** Based on the combined insights from the five analysis tasks, the SUS and the interviews, we compiled and rated a list of found issues inspired by Nielsen's severity ratings (Nielsen, 1993). This list includes "feature requests" (FR) and "severities" (SE) (see Table 3). All of these adjustments were included before the following focus group meetings.

**Table 3 – List of identified feature requests and severities (FR: 1: = nice to have, 2: = good feature, 3: = enhances usability; SE: 1: = minor, 2: = big, 3: = disaster; Effort: 1: = min, 2: = average, 3: = max).**

| Description | FR | SE | Effort |
| --- | --- | --- | --- |
| KDB: Include number of underlying elements | 2 | - | 3 |
| KDB: Gray out inactive elements | 2 | - | 2 |
| KDB: Automatically unfold by hover | 3 | - | 3 |
| Filter: Adding the number of elements in relation to the knowledge state | 2 | - | 1 |
| Filter: Consistency of interface labels | - | 2 | 1 |
| Connection line: Selected rule has to come back into focus by click | 2 | - | 2 |
| Arc-diagram: Reduce number of arcs | - | 2 | 1 |
| Tables: Change "Rule" to "Calls in Rule" | - | 2 | 1 |
| Tables: Change "=" to " = =" (for equal distribution) | - | 2 | 1 |
| … | | | |

### 6.3. Industry focus group

After the most important adjustments (see Table 3) were carried out in the prototype, we conducted two focus group meetings to get feedback on the changes and gather additional suggestions for improvement. Members of two different professional companies in the IT-security domain participated in these sessions. Each focus group contained three IT-security experts from industry.

**Design and procedure:** The focus group meetings were designed as guided meetings – we showed the prototype to the group and explained the work steps for analysis tasks. Each group member was asked to express feedback and provide suggestions at any time.

**Apparatus and material:** As material, we used a fully functional build of the KAMAS prototype including all improvements based on prior evaluation results. The guided focus group meetings were performed on a 15″ notebook with HD screen resolution and an external mouse. Both focus group meetings were held in early November 2015, with a duration of approximately one hour and documented on paper by the experiment facilitator.

**Results:** All focus group members reported that the prototype is well designed and very interesting to use because it gives each user the ability to benefit from the expert knowledge of others. One group member noted that it is hard to build a mental connection between the histogram and the related table columns. He recommended changing the histogram's background to the color of the related bars in the columns. One of the major inputs of the industry focus groups was that not only malignant rules, but also benign rules are important for good analysis result. Additionally, one company offered to carry out a long-term field study and to expand the prototype for their requirements.

## 7. Reflection and conclusion

Following the design study methodology (Sedlmair et al., 2012), reflection is the third contribution to a design study (retrospective analysis and lessons learned) for the improvement of current guidelines. When we broke down the reflection on our requirements from Section 2, the different validation steps confirmed that our interactive visualization prototype fulfills the requirements of malware analysts.

*R1 Data:* The cluster-grammar provided by IT-security experts is a complex data structure consisting of the derivation rules, detailed derivation information and occurrences. Using this information as basis, we designed three analysis tables for representing the data: 1) a call overview table, visualizing all calls included in the cluster-grammar combined with their total occurrence; 2) the rule overview table to represent the calls included in a rule by a graphical summary; to provide the comparability of rules in combination with the total occurrence of the rule, its length and derivation information; 3) a rule detail table, showing the included calls of a rule in their sequential order. To gain better insights into the rules distribution in relation to their occurrence and length, KAMAS provides two different histograms, each combined with a range slider which can be used as filtering option.

*R2 Visual representation:* In general, the decision for an interface similar to programming IDEs was well received by the participants. It was easy for them to understand the handling and to work with it. Additionally, the participants and the focus group members appreciated the prototype's wide range of coordinated features, which they regarded as useful for data exploration while not being overloaded. A particularly interesting outcome of the tests from a visualization design perspective is that the arc-diagrams did not provide the benefits we expected (e.g., easily locate and compare patterns in the call sequences). One participant realized that something interesting was in the data, but he could not pinpoint the meaning. Yet, the simple connection line between the rule overview table and the rule detail table, which originally was considered "nice to have" by designers, turned out to be a much appreciated and valuable feature. Thus, the connection line supports the analysts in finding the connection between these two representations.

*R3 Workflow:* All included filter methods were very well received. Additionally, the dynamic query concept and its fast response was understood very well by the participants. In general, they described the relationships of the filters as intuitive and the usage of the KDB by drag and drop actions as easy to use. By subsequent focus group meetings, further improvements were integrated, such as the colored background of the filters, used to emphasize visualization elements and connections. One of the major inputs of the industry focus group was that not only malignant rules, but also benign rules are important for good analysis. Therefore, participants suggested changing the three-step red color scale to a five-step scale with a neutral color in the center (see Fig. 8). Based on the insights we gained from the tests we found that for the participants, the visualization of the expert knowledge and also the handling of the KDB was easy to understand and use.

*R4 Expert knowledge:* As previously mentioned, the KDB tree was well received by the participants and focus groups members. Another improvement added as a result of the user study was the addition of brackets with numbers of included rules at the end of each node. Additionally, we added a counter for each type of represented knowledge in the interface (see Fig. 4.2.f). The industry focus group members noted that the newly added numbers were helpful for getting a better overview of the loaded data. Reflecting on the insights we gained in the performed tests, we found out that the analysts appreciated and could benefit from the externalized expert knowledge by sharing and activating or deactivating KDB elements during the analysis process.

**Categorization of KAMAS:** If we categorize KAMAS along the Malware Visualization Taxonomy (Wagner et al., 2015), we can see that KAMAS can be categorized on the one hand as a Malware Forensics tool which regards to the analysis of malicious software execution traces. On the other hand, KAMAS can also be categorized as a Malware Classification tool for malware comparison in relation to the automated analysis. This automated analysis works on the included call sequences contained in a loaded file and is based on the explicit knowledge stored in the KDB.

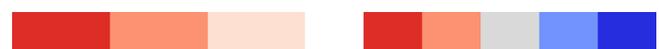

**Fig. 8** – Current three-step color map (left) versus a suggested five-step color map for knowledge representation (right). For color-blind users we decided to select to a diverging red-blue scale (red: = malicious, blue: = benign). (For interpretation of the references to color in this figure legend, the reader is referred to the web version of this article.)

**Lessons learned:** During this design study, we learned that explicit knowledge opens the possibility to close the gap between different categories of malware analysis systems (Wagner et al., 2015). Thus, it combines features for Malware Forensics and Malware Classification. In contrast to other malware analysis systems which build their visual metaphors directly on the output of the data providers, KAMAS uses an input grammar generated by a combination of Malheur (Rieck, 2016) and Sequitur (Nevill-Manning and Witten, 1997) for cluster and data classification. Therefore, we use analytical and visual representation methods to provide a scalable and problem-tailored visualization solution following the visual analytics agenda (Keim et al., 2010; Thomas and Cook, 2005). For keeping up with the large number and dynamic evolution of malware families, malware analysts need to continuously adapt the settings of their visualization systems, whereby interactivity is a key strength of visualization systems. Malware analysis in particular profits from extensive interaction and annotation features as it is a very knowledge-intensive job. By providing knowledge-oriented interactions, externalized knowledge can subsequently be used in the analysis process to improve analysts' performance.

**Transferability:** The knowledge generation loop can be generalized for other domains taking into account domain-specific data structures and patterns of interest. On a general level, the workflow for knowledge generation and extraction is mostly similar and always includes the system user as an integral part of the loop (Endert et al., 2014). Focusing on *n* stepped colored highlighting and easy to understand summarization techniques it is faster and more effective to find similarities in the data.

In future work, we will extend our research to a different problem domain (e.g., healthcare) in order to generalize our results. Therefore, we have to adapt and extend the knowledge-assisted visualization methods and the interface as necessary in relation to the users' needs.

## Acknowledgments

This work was supported by the Austrian Science Fund (FWF): P 25489-N23 via the KAVA-Time project. Many thanks to our collaboration partners and our study participants as well to Christina Niederer for her feedback to our manuscript and her support.

Markus Wagner is a research associate at the Institute of Creative\Media/Technologies, St. Poelten University of Applied Sciences (Austria) and a PhD student at the Faculty of Informatics at Vienna University of Technology. His research activities involve knowledge-assisted visual analytics methods for behavior-based malware analysis. Markus studied Game Engineering and Simulation at the University of Applied Sciences Technikum Wien where he received his MSc degree in 2013 and Industrial Simulation at St. Poelten University of Applied Sciences where he received his BSc degree in 2011.

Alexander Rind is a research associate at the Institute of Creative\Media/Technologies, St. Poelten University of Applied Sciences (Austria). His research activities involve knowledge-assisted visual analytics methods for behavior-based malware analysis, interaction in visual analytics, and visualization of electronic health records. Alexander studied Business Informatics in Vienna and Lund and received his MSc degree from Vienna University of Technology in 2004.

Niklas Thür is a student researcher at the Institute of Creative\Media/Technologies, St. Poelten University of Applied Sciences (Austria).

Wolfgang Aigner is a professor at St. Poelten University of Applied Sciences (Austria) and scientific director of the Institute of Creative\Media/Technologies. His main research interests include visual analytics and information visualization. Wolfgang Aigner received his PhD degree and habilitation from Vienna University of Technology in 2006 and 2013. He authored and co-authored more than 100 peer-reviewed articles as well as the book "Visualization of Time-Oriented Data" (Springer, 2011) that is devoted to a systematic view on this topic.